\documentclass[twocolumn,nodate,shownopacs,preprintnumbers,nofootinbib,amsmath,amssymb,aps,prl,superscriptaddress]{revtex4}

\usepackage{graphicx, wrapfig}
\usepackage{amssymb}
\usepackage{color}
\usepackage{mathrsfs}
\usepackage{amsmath}
\usepackage{subfigure}
\usepackage{afterpage}

\usepackage{epstopdf}

\DeclareGraphicsExtensions{.pdf,.png,.jpg,.jpeg,.eps}

\def\f{\frac}

\def\ig{\includegraphics}

\usepackage{enumerate}

\def\kz{k_{\circ}}

\def\Az{\mathring{A}}
\def\be{\nopagebreak[3]\begin{equation}}
\def\ee{\end{equation}}
\def\ba{\nopagebreak[3]\begin{eqnarray}}
\def\ea{\end{eqnarray}}

\newcommand{\bfig}{\begin{figure}[h]}
\newcommand{\efig}{\end{figure}}

\newcommand{\bmult}{\nopagebreak[3]\begin{multline}}
\newcommand{\emult}{\end{multline}}

\def\planck{{PLANCK} }

\begin{document}

\title{Alleviating the Tension in the Cosmic Microwave Background\\ using Planck-Scale Physics}

\author{Abhay Ashtekar}
\affiliation {
Institute for Gravitation and the Cosmos \& Physics Department,\\ 
Penn State University, University Park, Pennsylvania 16801, USA}
\author{Brajesh Gupt}
\affiliation {
Institute for Gravitation and the Cosmos \& Physics Department,\\ 
Penn State University, University Park, Pennsylvania 16801, USA}
\author{Donghui Jeong}
\affiliation{
Institute for Gravitation and the Cosmos \& Department of Astronomy and Astrophysics,\\
Penn State University, University Park, Pennsylvania 16801, USA}
\author{V. Sreenath}
\affiliation{
Department of Physics, National Institute of Technology Karnataka, Surathkal, India 575025}

\begin{abstract}
Certain anomalies in the CMB bring out a tension between the six-parameter flat $\Lambda$CDM model and the CMB data. We revisit the \planck analysis with loop quantum cosmology (LQC) predictions and show that LQC alleviates both the large-scale power anomaly and the tension in the lensing amplitude. These differences arise because, in LQC, the \emph{primordial} power spectrum is scale dependent for small $k$, with a specific power suppression. We conclude with a prediction of larger optical depth and  power suppression in the $B$-mode polarization power spectrum on large scales. 

\end{abstract}

\maketitle

\emph{Introduction.\textemdash}The $\Lambda$CDM model selected by the \planck satellite data has had impressive success in explaining all major features in the temperature anisotropies and polarizations of the cosmic microwave background (CMB), using only six parameters \cite{planck1}. Let us begin by recalling the procedure that is used to determine the model from the CMB data. Inspired by inflationary models, one assumes that the primordial power spectrum is nearly scale invariant with a specific form, which we will refer to as the standard ansatz (SA): 
\be \label{SA} 
\mathcal{P}_{\mathcal{R}}(k) = A_{s}\, 
\left(\f{k}{k_{\star}}\right)^{n_{s}-1}\!\! , \ee 
where $A_{s}$ is the amplitude of the scalar mode and $n_{s}$ its spectral index. (Here, $k_\star=0.05\,{\rm Mpc}^{-1}$ is the pivot mode.) To determine a specific $\Lambda$CDM model, one requires four additional parameters: $\Omega_{b}h^2,\, \Omega_{c}h^2$ that refer, respectively, to baryonic and the cold matter density; and $100\theta_*$, $\tau$ that determine the observed angular scale associated with acoustic oscillations, and the optical depth that characterizes the reionization epoch \cite{planck47}, respectively. Given the SA and the six parameters, the Boltzmann codes \cite{cmbfast,camb,class}  that incorporate subsequent astrophysics provides us with four power spectra $C_{\ell}^{TT}, \, C_{\ell}^{TE},\, C_{\ell}^{EE}, \, C_{\ell}^{\phi\phi}, \,$ where $T, E, \phi$ stand for temperature, $E$-mode (even-parity) polarization, and lensing potential \cite{planck5,planck6}.  One compares these theoretical predictions with the observed power spectra and finds the best-fitting values (together with uncertainties) for the six parameters. This fixes the $\Lambda$CDM model. One can then work out predictions for other observables, which can be  measured independently. For example, the four-point correlation function of the CMB measures the gravitational lensing amplitude $A_L$ \cite{Hu:2001kj}, and  the $B$-mode (odd-parity) polarization power spectrum $C_{\ell}^{BB}$ measures the amplitude of tensor perturbation in the early Universe  \cite{Kamionkowski:1996ks,Seljak:1996gy}.

At the same time, the CMB data exhibit some anomalies that bring out tensions between the best-fit $\Lambda$CDM model and observation. We will ignore the tension between the CMB and low-$z$ observations, and focus instead on two anomalies in the CMB.
The first is the large-scale power anomaly related to $S_{1/2} \equiv\int_{-1}^{1/2}\left[C(\theta)\right]^2d(\cos\theta)$, obtained by integrating the two-point correlation function $C(\theta)$ of the CMB temperature anisotropies over large angular scales ($\theta>60^\circ$). 
The WMAP \cite{wmap,sarkaretal} and \planck \cite{schwarzetal,planck7}  measured values of  $S_{1/2}$ are much smaller than the expectation from  the SA+$\Lambda$CDM cosmology.  The second is the anomaly associated with the lensing amplitude $A_{L}$. When it is allowed to vary, $A_L$  prefers a value larger than unity, hinting at an internal inconsistency in the $\Lambda$CDM cosmology  \cite{planck6,Motloch:2018pjy,Couchot:2015eea,Addison:2015wyg,Motloch:2019gux,Hadley} based on the SA. In particular,  it was recently suggested \cite{silketal} that this anomaly gives rise  to a ``possible crisis in cosmology''   because the positive spatial curvature one can introduce to alleviate this tension makes CMB analysis inconsistent with 
low-$z$ cosmological measurements.

In this Letter, we present an intriguing possibility of alleviating both 
anomalies within a well-motivated theoretical framework of loop quantum 
cosmology (LQC). First, the LQC prediction modifies the SA for the 
primordial power spectrum by suppressing its large-scale amplitude, which 
naturally leads to lower $S_{1/2}$. The scale-dependent primordial power 
spectrum, in turn, prefers a higher amplitude $A_s$ that pushes
lensing amplitude $A_L$  toward unity (making it consistent with flat
$\Lambda$CDM), and higher optical depth $\tau$. Finally we show
that, due to the modified primordial power spectrum and higher $\tau$, LQC leaves a
specific signature in the $B$-mode (odd-parity) polarization power spectrum. \vskip0.1cm

\emph{Modified primordial power spectrum.\textemdash}In LQC, the big bang singularity is naturally resolved and replaced by a big bounce (see, e.g., Refs. \cite{asrev,iapsrev}). Therefore, one can systematically investigate the dynamics of cosmological perturbations in the pre-inflationary epoch starting from the Planck regime  (see, e.g., Refs. \cite{aan1,aan3,madrid,bcgmrev,aaab,agullomorris,agulloassym,ag2,ag3,iabbvs,vsiabb,menaetal}). Since the quantum corrected Einstein's equations never break down, all physical quantities remain finite. In particular, while the scalar curvature  $R$  of space-time diverges at  the big bang,  it remains \emph{finite} at the bounce, achieving its universal maximum value $R_{\rm max} \approx 62$ in Planck units. Now, curvature\textemdash more precisely $R/6$\textemdash provides a natural scale in the dynamics of the gauge invariant perturbations  (which in de Sitter space-time coincides with  $2H^{2}$).  Fourier modes with \emph{physical}  wave numbers  $k_{\rm phys} \equiv k/a(\eta)\, \gg  (R/{6})^{1/2}$ are essentially unaffected by curvature while those with $k_{\rm phys} \lesssim  (R/{6})^{1/2}$ get excited. Therefore the evolution during the preinflationary epoch of LQC is subject to  a new scale: $k_{\rm LQC} = (R_{\rm max}/{6})^{1/2} \approx 3.21$ in Planck units. Modes with $k_{\rm phys}^{\rm B} \lesssim k_{\rm LQC}$ at the bounce are excited during their preinflationary evolution. Therefore they are not in the Bunch Davies (BD) vacuum at the onset of the relevant slow roll phase of inflation\textemdash i.e., a couple of $e$-folds before the time at which the mode with the largest observable wavelength crosses the Hubble horizon during inflation.  (For details, see Refs. \cite{aan1,aan3}).

Now, one's first reaction may be that these excitations are observationally irrelevant because they would be simply diluted away by the end of inflation. However, this is not the case: because of stimulated emission, the number density of these excitations remains constant during inflation \cite{ap,iapsrev,gk}.  Therefore the primordial LQC power spectrum at the  end of inflation is \emph{different} from the standard ansatz  of Eq.~(\ref{SA}) for modes with $k_{\rm phys}^{\rm B} < k_{\rm LQC}$.

The key question then is whether these long wavelength modes are in the observable range. The answer depends on the choice of the background metric that satisfies the quantum corrected Einstein's equations of LQC, and the Heisenberg state of the cosmological perturbations. In standard inflation, the background metric can be any solution of Einstein's equation for the given potential, and, since one cannot specify the quantum state of perturbations at the big bang, one specifies it, so to say, in the middle of the evolution by asking that they be in the BD vacuum at the start of the relevant phase of the slow roll. In LQC,  geometry is regular at the bounce.  Using this fact, key features of the quantum geometry in LQC, and a ``quantum generalization'' of Penrose's Weyl curvature hypothesis \cite{rp-weyl},  a specific proposal has been put forward to make the required choices \cite{ag2,ag3}.  Quantum corrected LQC dynamics then leads to unique predictions for the primordial power spectrum for any given inflationary potential; there are no parameters to adjust. The viewpoint is to use the proposal as a working hypothesis, analyze the consequences, and use the CMB observations to test its  admissibility. 

The proposal constrains the background metric to be such that the $\Lambda$CDM  universe has undergone  $\simeq \!141$ $e$-folds since the bounce (irrespective of the choice of inflationary potential) \cite{ag3}. It then follows that  the mode with $k_{\rm phys} = k_{\rm LQC}$ at the bounce has comoving wave number $\kz \simeq 3.6\times10^{-4}{\rm Mpc}^{-1}$. The primordial power spectrum of LQC is nearly scale invariant for $k \gg \kz$ but power is suppressed for $k\lesssim 10 \kz$: 
  \be \label{LQC} \mathcal{P}_{\mathcal{R}}^{LQC} 
 = f(k) \,A_{s}\, \left(\f{k}{k_{\star}}\right)^{n_{s}-1}\!\!\!\! ,\ee 
 where the form of the suppression factor $f(k)$ can be seen in
Fig.\ref{fig:supsupPower}.  [$f(k) \approx 1$ for $k \gg \kz$.] This difference
from the standard ansatz can be traced back directly to the modes not being in
the BD vacuum at the onset of inflation. Now, if the total energy in
the scalar field is dominated by the kinetic contribution at the bounce, details
of the potential do not affect the preinflationary dynamics, and the
suppression factor $f(k)$ is also the same. Analysis of Ref. \cite{Copelandetal} strongly suggests 
that there is a large class of potentials for which our proposal to choose the
background geometry will constrain the bounce to be kinetic energy dominated. This is illustrated by comparing the Starobinsky inflation \cite{Starobinsky:1980te} and the quadratic potential in Fig.\ref{fig:supsupPower}.

\bfig
\hskip-.5cm
 \ig[width=3.2in]{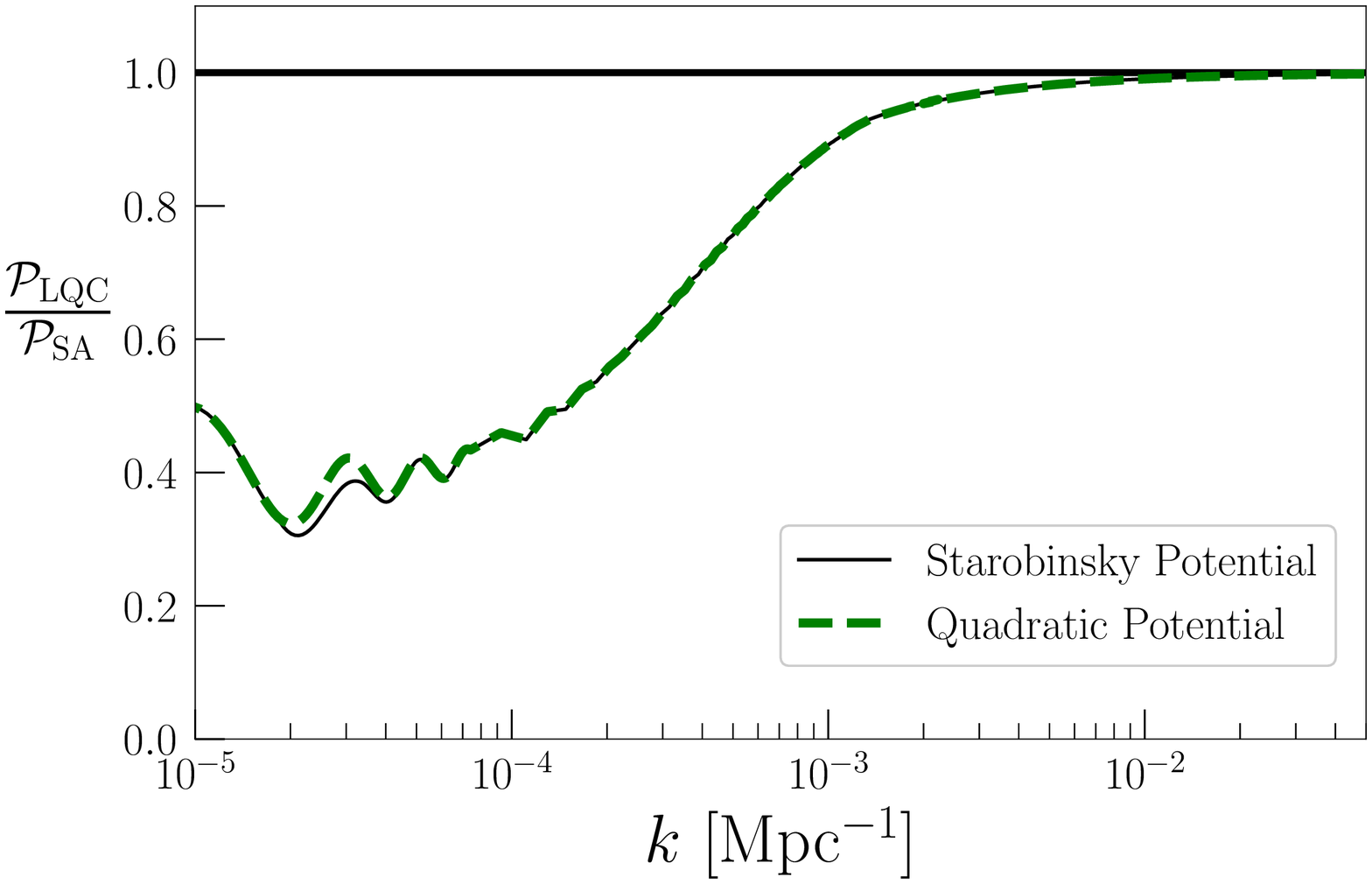}
  \caption{Ratio of the primordial scalar-power spectrum for LQC and SA.  Power is suppressed in LQC for $k\lesssim 10 \kz \simeq 3.6\times 10^{-3}{\rm Mpc}^{-1}$. Plots for the Starobinsky and quadratic potentials are essentially indistinguishable.} 
 \label{fig:supsupPower}
\efig
\bfig
\ig[width=3.2in]{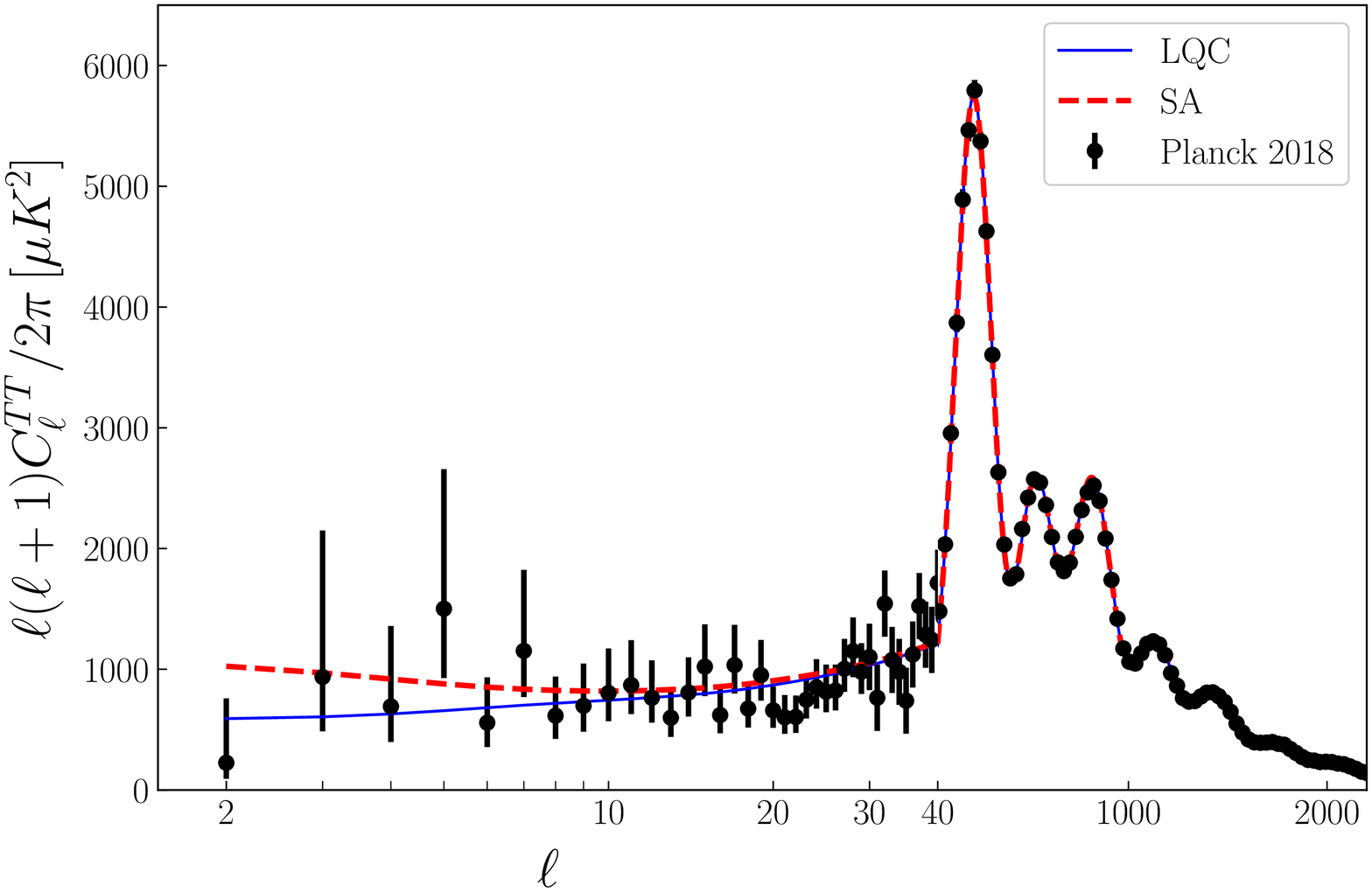}
 \caption{TT power spectra. The 2018 \planck spectrum (black dots with error bars),  the LQC [solid (blue) line], and the standard ansatz predictions [dashed (red) line].} 
\label{fig:TT}
\efig
\emph{Results.\textemdash}All results are based on the \planck\!\!-2018 data
\cite{planck1} using the observed TT, TE, EE, and $\phi$-$\phi$ power spectra
(including the $\ell <30$ modes for EE correlations) to which the associated
likelihoods are Planck TTTEEE+lowl+low$E$+Lensing.

Figure~\ref{fig:TT} shows the observed TT-power spectrum together with 
the 1$\sigma$ (68\% confidence level) error bars, and the LQC and the 
SA predictions for the respective best-fit cosmological parameters.
Clearly, LQC power is suppressed at $\ell\lesssim 30$ relative to the SA. This is also true for the EE power spectrum (as already noted in Ref. \cite{ag3}, using the then available \planck\! 2015 data).
Note that the difference between LQC and SA best-fitting curves shown in 
Fig.~\ref{fig:TT} underestimates the difference in the predicted 
primordial spectra, for the best-fitting cosmological parameters are different. 
Also, had the LQC+$\Lambda$CDM model been used for their analysis, the cosmic-variance uncertainties on large scales may have been 
smaller than the reported values from \planck\! 2018.

Figure~\ref{fig:Ctheta} compares the angular TT two-point correlation function
$C(\theta)$ predicted by LQC with that predicted by the SA. It is clear by
inspection that the LQC prediction for $C(\theta)$ is closer to the observed
values  for all $\theta$. In order to quantify this difference, we computed
$S_{1/2}$. As the last row of Table I shows, the $S_{1/2}$ from the best-fit
LQC+$\Lambda$CDM model is about \emph{a third} of that obtained from
SA+$\Lambda$CDM, and closer to the value of $S_{1/2}=1209.2$ given by the PLANCK Collaboration using the Commander CMB map. But since that value is obtained after masking and additional processing,  a more appropriate comparison would be with the value $6771.7$ of 
$S_{1/2}$ obtained from the full sky map, i.e. using the PLANCK $C^{TT}$ data for all $\ell$.  
The difference between LQC and this PLANCK value is also significantly lower than that between 
SA and this PLANCK value. This is a substantial alleviation of the tension between theory
and observations that has been emphasized over the years
\cite{wmap,sarkaretal,schwarzetal,planck7}.
\bfig
  \ig[width=3.2in]{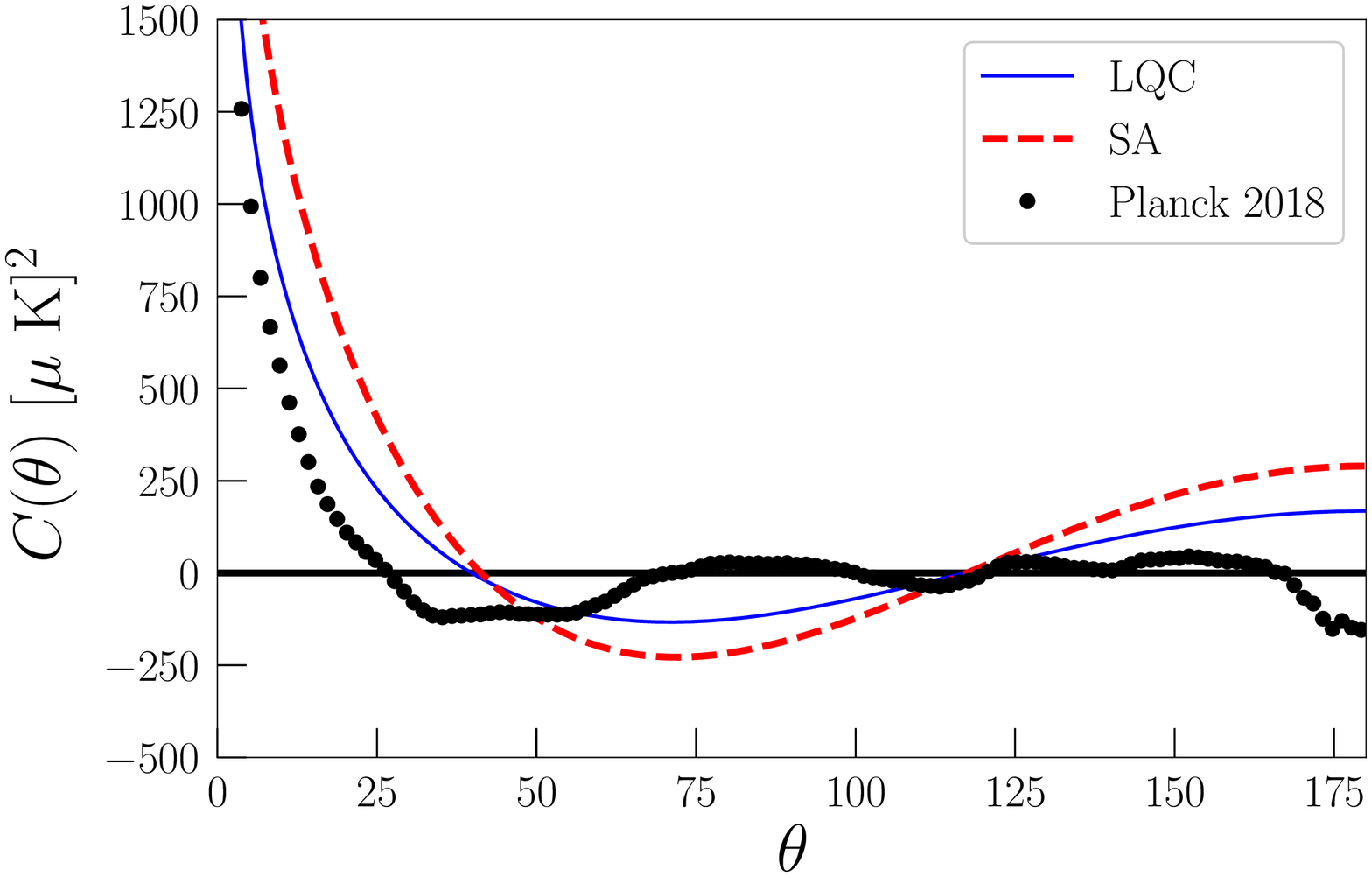}
 \caption{The angular power spectrum $C(\theta)$. The 2018 \planck spectrum (thick black dots),  the LQC  [solid (blue) line], and the standard ansatz [dashed (red) line] predictions.Values of cosmological parameters are fixed to the mean values given in Table \ref{table1}.}  
 \label{fig:Ctheta}
\efig
\begin{table}
\begin{center}
\footnotesize
\begin{tabular}{| c | c | c | }
\hline
{\rm Parameter}   &                         {\rm SA}               &                                               \rm{LQC}
 \\ \hline \hline

$\Omega_b h^2$  &     $0.02238 \pm 0.00014$       &                           $0.02239 \pm 0.00015$ \\ \hline
$\Omega_c h^{2}$    &     $0.1200\pm 0.0012$         &                         $0.1200\pm0.0012$ \\ \hline
$100\theta_{MC}$        &              $1.04091 \pm 0.00031$          &                       $1.04093 \pm 0.00031$ \\  \hline
$\tau$           &              $0.0542\pm 0.0074$             &                     $0.0595\pm0.0079$  \\ \hline
$\ln (10^{10} A_s)$   &    $3.044\pm 0.014 $         &                        $3.054\pm0.015$ \\ \hline
$n_s$               &          $0.9651 \pm 0.0041$               &                      $0.9643\pm0.0042$   \\ \hline
\hline
$S_{1/2}$           &          $ 42496.5$                                                   &               $14308.05$
\\   \hline
\end{tabular}
\caption{Comparison between the standard ansatz and LQC.  The mean values of the marginalized PDF for the six cosmological parameters, and values of $S_{1/2}$ calculated using $C_{\ell}^{\rm TT}$.}
\label{table1}
\end{center}
\end{table}

Table I also shows the mean values of the marginalized probability distributions of the six cosmological parameters together with their $1\sigma$ ranges. 
For the first three, namely, $\Omega_b h^2$, $\Omega_c h^2$, and $100\theta_{MC}$,
the difference between the SA+$\Lambda$CDM and LQC+$\Lambda$CDM values is $<
0.07\sigma$ and for $n_s$ the difference is $\sim0.2\sigma$. 
However, the values of the optical depth $\tau$ and $\ln (10^{10} A_s)$ have increased in
LQC by $0.72\sigma$. As we discuss below, this significant change is a
direct consequence of the scale-dependent initial power spectrum
(\ref{LQC}) of LQC, which also leads to a $0.56\sigma$ decrease in the lensing
amplitude $A_{L}$ from $1.072\pm0.041$ in SA+$\Lambda$CDM to $1.049\pm0.040$ in
LQC+$\Lambda$CDM, when $A_L$ is also varied. Furthermore, when $A_L$ is included in the 
analysis, the $\Lambda$CDM parameters change by $0.59\sigma-1.48\sigma$ 
in SA and $0.39\sigma-1\sigma$ in LQC.
As Fig.~\ref{fig:taual} shows, the value  $A_{L} \!=\!1$
lies outside of the 68\% confidence level for the SA+$\Lambda$CDM model (red
contours).  A natural way to alleviate this tension within the SA+$\Lambda$CDM
is to consider a closed universe. However, then other disagreements with
observations arise that prompted the authors of Ref. \cite{silketal} to raise the
possibility of a ``crisis in cosmology.'' What is the situation with the altered
values of $\tau$ and $A_{L}$ in LQC? We see from  Fig.~\ref{fig:taual} that now
the tension is naturally alleviated because the value $A_{L}\! =\!1$ is
within 68\% confidence level (blue contours). Therefore, the primary motivation
for introducing spatial curvature no longer exists  in LQC. \vskip0.1cm
\bfig
 \ig[width=3.2in]{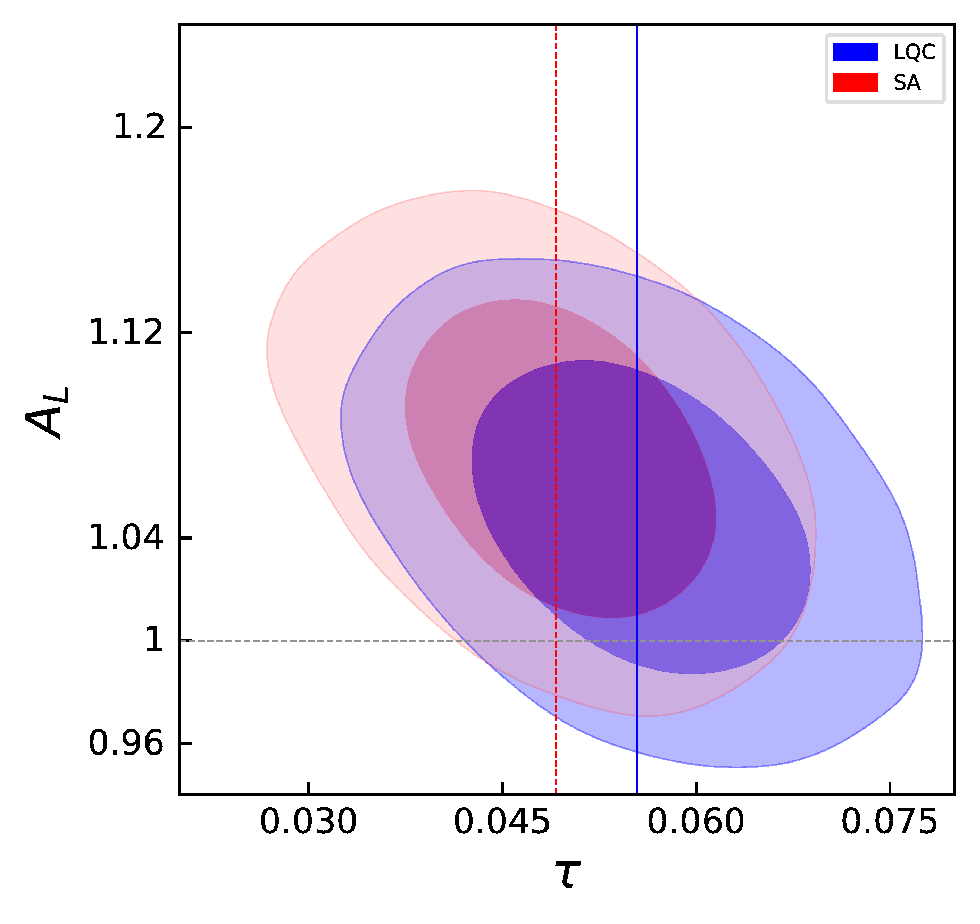}
 \caption{$1\sigma$ and $2\sigma$ probability distributions in the $\tau\!-\!A_{L}$ plane. Predictions of the standard ansatz (shown in red) and LQC (shown in blue). Vertical lines represent the respective mean values of $\tau$.} 
\label{fig:taual}
\efig

\emph{General implications of power suppression at large angles.\textemdash}In LQC, the mechanism for departure from the nearly scale invariant ansatz (\ref{SA})  is rooted in fundamental considerations in the Planck regime. Nonetheless, it is natural to ask if the qualitative features of some of our results will carry over if there were other mechanisms that led to the primordial spectrum of the form given in Eq.~(\ref{LQC}). We now show that this is indeed the case. 

Let us then suppose that there is \emph{some} mechanism that provides a  primordial power spectrum of the form (\ref{LQC}) for some $\kz$.  Let us compare and contrast the resulting best fit $\Lambda$CDM model with that given by the SA of Eq.~(\ref{SA}). As a first step, let  us restrict the analysis only to smaller angular scales ($k\gg \kz$). Then, the primordial spectrum in both schemes is the same, whence we  will obtain the same best fit values of the six cosmological parameters.  Denote by $\Az_{s}$ the best fit value of the scalar amplitude $A_{s}$.  In the second step, let us bring in the \emph{full} range of observable modes including $k \le \kz$. Now,  given the  \emph{observed} large-scale suppression in the  TT power spectrum,  for SA+$\Lambda$CDM model the best-fit value $A_{s}^{\rm (1)}$ for the entire $k$ range will be lower than $\Az_{s}$. By contrast if the primordial power spectrum is of the form of Eq.~(\ref{LQC}), $\Az_{s}$ will not have to be lowered as much to obtain the best fit $A_{s}^{(2)}$ since the initial power is already suppressed by $f(k)$. Thus, we have $\Az_{s} > A_{s}^{(2)} > A_{s}^{(1)}$.  [For the $f(k)$ in LQC, we have $\ln (10^{10}\Az_{s}) = 3.089$ and  $\ln (10^{10} A{s}^{(2)}) = 3.054$ and $\ln (10^{10}A{s}^{(1)}) = 3.044$.]  The key point is the last inequality: $A_{s}^{(2)} > A_{s}^{(1)}$.  Now, we know that for large $k$, the product $A_{s}e^{-2\tau}$ is fixed  by observations. Hence, it follows that the best fit values of the optical depth in the two scheme must satisfy $\tau^{(2)} > \tau^{(1)}$. Finally, from the very definition of lensing amplitude, the value of  $A_{L}$ is anticorrelated to the value of $A_{s}$. Therefore, we will have $A_{L}^{(2)} < A_{L}^{(1)}$. Thus  in any theory that has  primordial spectrum of the form  (\ref{LQC}),  $A_{s}, \tau$, and $A_{L}$ will have the same qualitative behavior as in LQC, and hence the tension with observations would be reduced.
What LQC provides is a precise form of the suppression factor $f(k)$ from ``first principles,'' and hence specific quantitative predictions. The LQC $f(k)$ also leads to other predictions\textemdash e.g., for the BB power spectrum discussed below\textemdash that need not be shared by other mechanisms.

\emph{Summary and discussion.\textemdash}In LQC, curvature never diverges and reaches its maximum value at the bounce. As a result,  preinflationary dynamics naturally inherits a new scale, $k_{\rm LQC}$, such that modes with $k_{\rm phys}^{B} \lesssim k_{\rm LQC}$ at the bounce are not in the BD vacuum at the start of the slow roll phase of inflation \cite{aan1,aan3}, whence the primordial power spectrum is no longer nearly scale invariant, but of the form (\ref{LQC}). The LQC dynamics and initial conditions then imply \cite{ag3} that there is power suppression in CMB at the largest angular scales $\ell \lesssim 30$. In contrast to other mechanisms that have been proposed, this suppression has origin in fundamental, Planck
scale physics rather than in phenomenological adjustments put in by hand just
before or during the slow roll. As a result of this power suppression, there is an enhancement of optical depth $\tau$ and suppression of the lensing
potential $A_{L}$. The two together bring the value $A_{L}\!=\!1$ within
$1\sigma$ of the LQC $\tau\!-\!A_{L}$ probability distribution, thereby removing
the primary motivation for considering closed universe and the subsequent
``potential crisis'' \cite{silketal}. In addition, the anomaly in
$C(\theta)$ at large angles \cite{wmap,sarkaretal,schwarzetal,planck7}
is significantly reduced; the LQC value of $S_{1/2}$ is $\sim0.34$ of that
predicted by standard inflation. The \planck Collaboration had suggested
\cite{planck1} that  ``\ldots if any of the anomalies have primordial origin,
then their large scale nature would suggest an explanation rooted in fundamental
physics. Thus it is worth exploring any models that might explain an anomaly
(even better, multiple anomalies) naturally, or with very few parameters.'' In
this Letter we presented a concrete realization of this idea. (For an alternate
proposal within LQC see Ref. \cite{ivan2020}).
\bfig
\ig[width=3.2in]{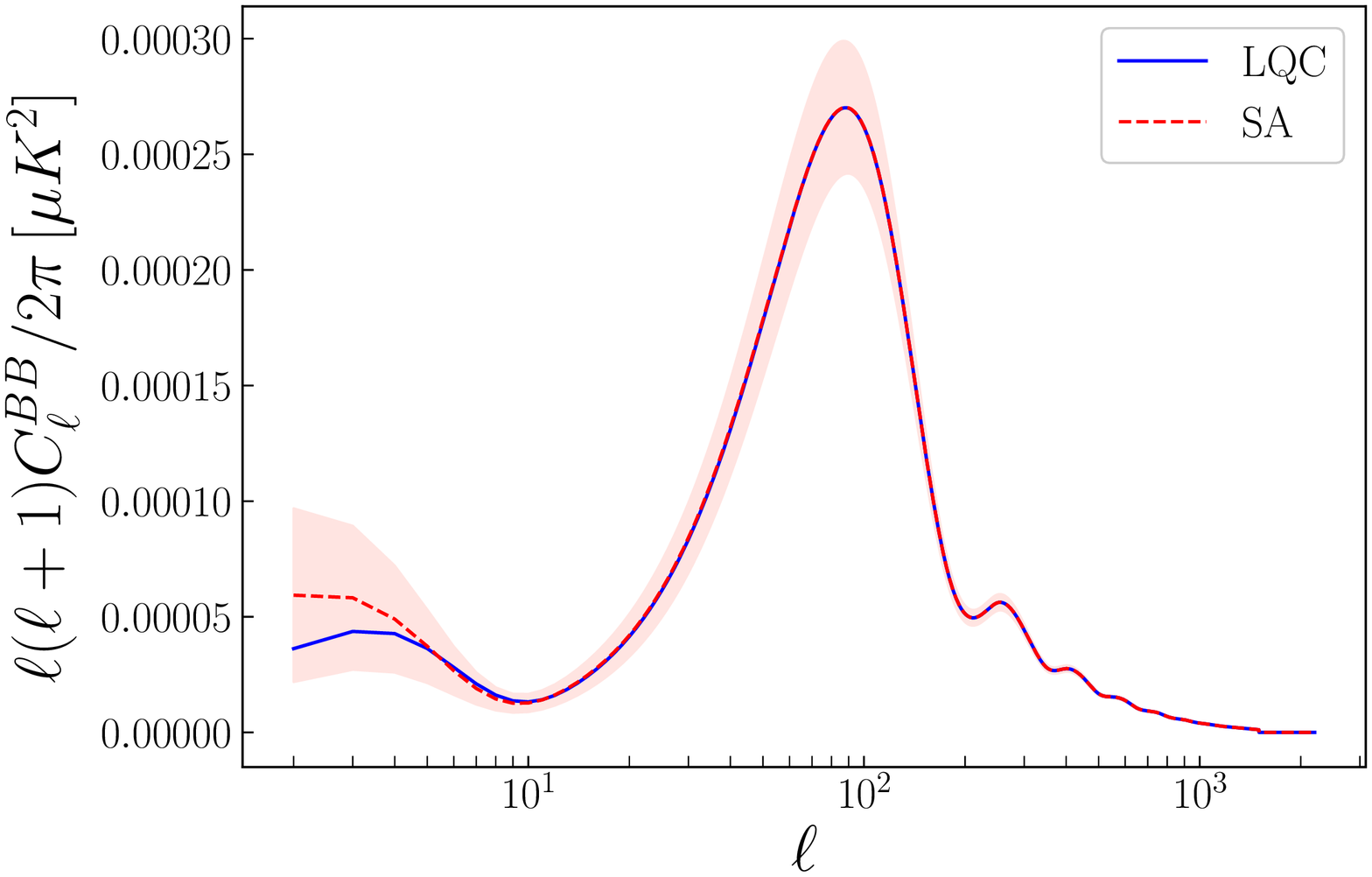}
 \caption{Predicted Power spectra for BB polarization with $1\sigma$ uncertainty. Comparison between LQC and standard inflation.  The tensor to scalar ratio $r$ has been set to $0.0041$, motivated by  Starobinsky inflation \cite{Starobinsky:1980te}. The shaded region indicates the cosmic variance for SA.}
 \label{fig:clbb}
\efig

This  model also leads to other specific predictions. First, as Table I shows, the reionization optical depth $\tau$ is predicted to be $\sim 9.8\%$ (i.e., $0.72\,\sigma$) higher. This prediction can be tested by the future observation of global 21 cm evolution at high redshifts that can reach a percent level accuracy in the measurement of $\tau$ \cite{Fialkov:2016zne}. Second, for any given inflationary potential,  the \emph{primordial} spectra of LQC and SA share the same value of  $r$\textemdash the tensor to scalar ratio\textemdash which depends on the potential. But there is a specific scale dependence in the large-scale $B$-mode (odd-parity) polarization power spectrum, as shown in Fig.~\ref{fig:clbb}. The difference is driven by the LQC suppression of the primordial tensor amplitude combined with the larger reionization contribution due to higher $\tau$. Provided that  $r$ is sufficiently large,  for example, $r\gtrsim 0.001$, we may be able to test this prediction against the data from the future $B$-mode missions
such as 
LiteBIRD \cite{Matsumura:2013aja}, 
Cosmic Origins Explorer \cite{core}, or Probe Inflation and Cosmic Origins (PICO \cite{Hanany:2019lle}).
Again, LQC  modifies $C_\ell^{BB}$ on large scales where the 
cosmic variance limits its detectability. However, in light of results presented in this Letter, we hope that the LQC primordial power spectrum will be included in the future cosmological analysis.

We would like to thank Ivan Agullo and Charles Lawrence for valuable inputs and Pawel Bielewicz  for help with Fig. \ref{fig:Ctheta}. This work was supported in part by the NSF Grants No. PHY-1505411 and No. PHY-1806356,\, the NASA ATP Grant No.\ 80NSSC18K1103, and the Eberly research funds of Penn State. V.S. was supported by the Inter-University Centre for Astronomy and Astrophysics, Pune during the early stages of this work. Portions of this research was conducted with high performance computing resources provided by Louisiana State University \cite{hpclsu}.

\end{document}